
\input harvmac

 \hoffset-0.2in

\Title{\vbox{\baselineskip12pt\hbox{FTUAM-35-93}}}
{\vbox{\centerline{Gauge Coupling Unification:}
   \vskip2pt\centerline{Strings versus SUSY-GUTs}}}

\centerline{Luis E. Ib\'a\~nez}
    \centerline{Departamento de F\'isica Te\'orica  C-XI}
\centerline{Universidad Aut\'onoma de Madrid}
\centerline{Cantoblanco, 28049 Madrid, Spain}

\vskip .2in
{\vbox{\centerline{ABSTRACT}}}

{}.

 \vskip .2in
 \noindent
\vbox{\baselineskip12pt
Standard (level one) heterotic string models with standard model gauge
group predict the unification of $SU(3)$ and $SU(2)$ gauge
couplings whereas the $U(1)$ factor is unified modulo an
unknown normalization factor $k_1$. On the other hand the
unification mass is known. I argue that this situation is
quite analogous (though opposite) to that in SUSY-GUTs in which
the $U(1)$ normalization is known ($k_1=5/3$) but the
unification mass $M_X$ is unknown. I emphasize that $k_1$
should be taken as a free parameter in the string
approach (quite in the same way as $M_X$ is taken as a
free parameter in SUSY-GUTs). If this is done, the success of
the string approach concerning gauge coupling unification
is comparable to that in SUSY-GUTs}

 \Date{August 1993}
\noblackbox

 \lref\GS{M. Green and J. Schwarz, Phys.Lett.B149 (1984) 117.}
\lref\LNS{E. Witten, Phys.Lett.B149 (1984) 351;
W. Lerche, B. Nilsson and A.N. Schellekens, Nucl.Phys.B299 (1988)
91;  M. Dine, N. Seiberg and E. Witten, Nucl.Phys.B289 (1987)
585; J. Atick, L. Dixon and A. Sen, Nucl.Phys.B292 (1987) 109.}
\lref\GQW{H. Georgi, H.R. Quinn and S. Weinberg, Phys. Rev.
Lett. ${\underline{33}}$ (1974) 451.}
\lref\GINS{P. Ginsparg, Phys.Lett.B197 (1987) 139.}
\lref\DRW{S. Dimopoulos, S. Raby and F. Wilczek, Phys.Rev.D24 (1981)
1681; L.E. Ib\'a\~nez and G.G. Ross, Phys.Lett.105B (1981) 439;
S.Dimopoulos and H. Georgi, Nucl.Phys.B193 (1982) 475.}
\lref\AMAL{J. Ellis, S. Kelley and D.V. Nanopoulos, Phys.Lett.249B
(1990) 441; Phys.Lett.260B (1991) 131; P. Langacker and M. Luo,
Phys.Rev.D44 (1991) 817; U. Amaldi, W. de Boer and H. Furstenau,
Phys.Lett.B260 (1991) 447}
\lref\FIQ{ A. Font, L.E. Ib\'a\~nez and F. Quevedo, Nucl.Phys.
B345 (1990) 389.}
\lref\ILR{L.E. Ib\'a\~nez, D. L\"ust and G.G. Ross, Phys.Lett.
B272 (1991) 251; L.E. Ib\'a\~nez and D. L\"ust, Nucl.Phys.B382
(1992) 305.}
\lref\AEKN{I. Antoniadis, J. Ellis, S. Kelley and D. Nanopouolos,
Phys.Lett.B271 (1991) 31.}
\lref\POL{P. Langacker and N. Polonsky, Phys.Rev.D47 (1993) 4028.}
\lref\KAPLU{V. Kaplunovsky, Nucl.Phys. B307 (1988) 145.}
\lref\THRES{L. Dixon, V. Kaplunovsky and J. Louis, Nucl.Phys.B355
(1991) 649; J.P. Derendinger, S. Ferrara, C. Kounnas and F. Zwirner,
Nucl.Phys.B372 (1992) 145.}
\lref\SCHEL{A.N. Schellekens, Phys.Lett.B237 (1990) 363.}
\lref\FENO{A. Font, L.E. Ib\'a\~nez, F. Quevedo and A. Sierra,
Nucl.Phys. B331 (1990) 421.}
\lref\I{L.E. Ib\'a\~nez, Phys.Lett.B303 (1993) 55.}
\lref\CM{A. Casas and C.Mu\~noz, Phys.Lett.B214 (1988) 543.}

In the last couple of years there has been a revival of interest
on the topic of gauge coupling unification at very high energies
\GQW .
More precise experimental data on gauge couplings has allowed
for a more detailed check of the predictions of Grand Unified
Theories (GUTs) for gauge couplings. In particular, it has been
confirmed that there is very good agreement of the results
predicted \DRW\ by supersymmetric (SUSY) unified theories with the
experimental results \AMAL . On the other hand, the predictions
of non-supersymmetric GUTs are many standard deviations away
from these experimental results. Thus, e.g., a
non-SUSY version of minimal $SU(5)$ is experimentally ruled
out by these measurements.

Given this spectacular success of the SUSY-GUT scenario
one is naturally lead to look closer to these theories
and see what is their theoretical status. There are two major
problems in these theories. The first of them is the
notorious {\it doublet-triplet splitting problem} of GUTs
(and SUSY-GUTs). In all GUT models the Higgs-doublet of the
standard model has colour-triplet GUT-partners. These
colour triplets have quantum numbers of $d$-quarks and
can mediate fast proton decay unless they have masses of the
order of the unification scale. Thus we have to split the
Higgs multiplet allowing for the usual light Higgs doublets
but giving a large mass to the coloured triplet fields.
Of corse, this can be done by fine-tuning but this would
bring us back a new (though milder) form of
gauge hyerarchy problem. Other mechanisms propossed to
perform this splitting without fine-tuning either do not
work or require quite baroque and unexpected (huge) new
multiplets. Thus it looks like if the Higgs sector
refused to be unified, although, on the other hand
its presence is essential in getting the appropriate
gauge coupling unification!

The other major problem of SUSY-GUTs is the difficulty in
unifying this type of theories with gravity in the context
of the only consistent known gravity-unified theories, strings.
Indeed, it is well known that in order to obtain a usual GUT
like $SU(5), SO(10)$ etc. from strings one needs to go
beyond the usual compactifications and consider theories
with the gauge sector
involving higher level Kac-Moody algebras.
Only these theories allow for adjoint superfields in
the massless level of the string. Although in
principle this is not necessarily a problem, these theories
are quite cumbersome in practice (see e.g. ref.\FIQ ).
Furthermore, once one has
succeed in constructing e.g. an $SU(5)$ string, it is very difficult
to avoid the presence of extra unwanted adjoints (or other
exotic multiplets like $15$s, $40$s, etc). On top of that,
pretending that all the Higgs multiplets in the model
have precisely the couplings required to perform the doublet-triplet
splitting is just hoping for a miracle. Thus the situation
concerning SUSY-GUTs is quite puzzling: they are theoretically
in quite a bad shape but give the correct result required for
unification!

In the present note I remark that there is another class of
well motivated theories which have comparable success
concerning gauge coupling unification but are free from the
theoretical problems of SUSY-GUTs mentioned above.
This class of theories correspond to the assumption that
the SUSY standard model is directly unified into a string
theory close to the Planck mass, without any GUT intermediate
step. In fact gauge coupling unification within this type of
scheme has been considered in the recent past reaching
apparently a different conclusion \ILR ,\AEKN . The origin of this
 difference is a matter of
 appropriately identifying what are actually
the free parameters in string unification. This will be
clarified  below.

Let us first briefly recall the situation in SUSY-GUTs. Here the
normalization  of the $U(1)$ factor is known ($k_1=5/3$) and the
one-loop expressions for the weak angle and $\alpha _s$ yield
\eqn\sgut{
\eqalign{
sin^2\theta _W(M_Z)\ =&\ {3\over 8}(1\ +\ {{5\alpha (M_Z)}\over
{6\pi }}\ (b_2-{3\over 5}b_1)\ log({{M_X}\over {M_Z}})\ )  \cr
{1\over {\alpha _s(M_Z)}}\ =&\ {3\over 8}({1\over {\alpha (M_Z)}}\ -\
{1\over {2\pi }}\ (b_1+b_2-{8\over 3}b_3)\ log({{M_X}\over {M_Z}})\ )  \cr
.}}
where one has $b_1=11, b_2=1$ and $b_3=-3$ in the SUSY case.
In principle $M_X$ is unknown and the formulae \sgut \ give us
a constraint between the values of $sin\theta _W$ and $\alpha _s$
consistent with unification. This constraint is shown numerically in the
figure in which it is represented as a line in the $sin\theta _W$-
$\alpha _s$ plane. Different points in the line correspond to
different values for $M_X$ ($log_{10}M_X$ is shown at various points
on the line).
We will not atempt here to include a detail treatment of the errors.
We have included an error band  corresponding to
an uncertaintity of $\pm 0.01$ in the resulting value for $\alpha _s$. There
are different sources of errors coming from the uncertainty in the
low-energy and superheavy thresholds, two-loop effects etc. (see
e.g., ref. \POL \ for a detailed discussion of these points).
The success of the SUSY-GUT predictions correspond to this line
going through the experimental results also depicted in the figure
($\alpha _s$ is taken from jet event shape analysis). On the other
hand the lower curve in the figure corresponds to the non-supersymmetric
GUT result. It is clear this latter case is ruled out.

Let us go now to the supersymmetric string case. We will consider
a situation in which the gauge group is of the form
$SU(3)\times SU(2)\times U(1)_Y\times G$, $G$ being some other
possible gauge group factors. Furthermore we will assume that
the only low-energy particles which are charged under the
standard model group are those of the minimal SUSY standard model.
The gauge coupling constants $g_i$ of the three $SM$ interactions
are related at the string scale by \GINS\
\eqn\strg{
k_3g_3^2\ =\ k_2g_2^2\ =\ k_1g_1^2
}
In any standard compactification of the $E_8\times E_8$ heterotic
string one has in fact $k_2=k_3=1$. And hence we have the boundary
condition $g_2^2=g_3^2$ at the string scale. Notice that this
boundary condition is present without any GUT type of symmetry
relating strong to weak interactions, is just a consequence of
direct (level one) string unification. The case of $k_1$ is different
and we only know that it is in general a fractional number
with $k_1\ge 1$ (see below for a brief discussion of our
limited knowledge about the possible values of $k_1$)).
Thus it looks one has less predictivity compared to the GUT case
since in the latter one has one more boundary condition
($g_3^2=g_2^2=5/3g_1^2$). In fact the predictivity is the same
because in the string case we also know the unification mass
$M_X$. It is related in a calculable manner to the Planck mass
and one finds \KAPLU\  $M_{string}=5.3g10^{17}$ GeV. Thus the fact that
strings are theories which are unified with gravity provide us
with an extra constraint not present in GUTs. Now, the
one-loop results for the weak angle and $\alpha _s$ yield
\eqn\strgg{
\eqalign{
sin^2\theta _W(M_Z)\ =&\ {1\over {1+k_1}}(1\ +\ {{k_1\alpha (M_Z)}\over
{2\pi }}\ (b_2-{1\over {k_1}}b_1)\ log({{M_{string}}\over {M_Z}})\ )  \cr
{1\over {\alpha _s(M_Z)}}\ =&\ {1\over {1+k_1}}({1\over {\alpha (M_Z)}}\ -\
{1\over {2\pi }}\ (b_1+b_2-(1+k_1) b_3)\
log({{M_{string}}\over {M_Z}})\ )  \cr
.}
}
In analogy with the SUSY-GUT case, these two expressions give a
constraint between $sin^2\theta _W$ and $\alpha _s$. Now the free
parameter is $k_1$ (instead of $M_X$) and this constraint
may be represented as a line in the  $sin^2\theta _W$-$\alpha _s$
plane, the upper line (band) in the figure. Different points
in the line correspond to different values for $k_1$ from
1.3 to 1.7 (some sample values for $k_1$ are shown on the line).
As in the GUT case we have included an estimated band of error.
Some of the error sources (low-energy threshold, two-loop effects)
are identical to those present in GUTs. The treatment of the
heavy threshold errors is different and can be substantially larger
in the string compared to the GUT case, as has been recently
emphasized. To give an idea of the uncertainties I just include, as in
the previous case, a band of error corresponding to an
uncertainty of $\pm 0.01$ in $\alpha _s$
(I briefly come back to the issue of string threshold
corrections  below). The figure shows that the band corresponding
to string unification is quite close to the experimental results
and hence one concludes that {\it the data is compatible with
direct string unification} at the string scale $M_{string}=5.3g10^{17}$
GeV. This is the simple fact I want to emphasize in this note.
The result is very strongly dependent on the value of $k_1$
and the best agreement is found for values $k_1\simeq 1.4$.
Notice also that the string case seems to prefer higher values
 of $\alpha _s(M_Z)$
compared to the GUT case.
Taking $k_1$ as a free parameter in gauge coupling unification
was previously considered in refs.\CM ,\FIQ and \I .

 There is a forth logical possibility
which is considering non-supersymmetric string unification. In
this case the results would be incompatible with data for any $k_1$:
one obtains e.g. $\alpha _s(M_Z)\simeq 1.0$. This shows that
the fact that in the supersymmetric string case agreement is found
is a non trivial result.

Let us end this note by adding a couple of comments about
threshold effects and a discussion of possible $k_1$ values
in string models.

A comment concerning the size of
string threshold corrections is in order. The form of these threshold
corrections has been computed \KAPLU ,\THRES\
for a class of 4-dimensional strings
	(orbifold compactifications) and it has been found that they grow
linearly with $R^2$, $R$ being the compactification radius of the
orbifold. This can potentially lead to large corrections.
En ref.\ILR\ it was found that one can achieve
consistent gauge coupling unification even taking $k_1=5/3$ for
a) sufficiently large $R^2\simeq 10-16$ b) if the matter fields
have appropriate transformation properties. The latter is not
always possible and this can lead to interesting constraints
in that class of models. In this approach
the corrections to $\alpha _s$ can be enormous, e.g.,of order
$\Delta \alpha _s\simeq -0.07$, as required to get agreement with
$k_1=5/3$ (see the point $k_1=5/3$ in the figure).
Concerning the value of $R^2$, such  relatively large values
would correspond to the compactification
scale being slightly below the string scale. The dynamical models
of supersymmetry breaking constructed up to now seem to prefer on the other
hand relatively low values for $R^2$ in the range $R^2\simeq 1.-3.$, as
a reflection of the $R\rightarrow 1/R$ duality symmetry of
toroidal compactifications. If one takes values $R^2\simeq 2-3$ one
finds typical shifts in $\alpha _s$ of order $\pm 0.01$, similar
to the error band shown in the figure. Thus the width of the band
in the figure just corresponds to assuming modest string threshold
corrections of this order of magnitude.

The second comment concerns the possible values of $k_1$ in
string models.  Not much is known about $k_1$ , apart from the fact that it
should be a rational number. A massless state which transforms
like the representation $(R_3,R_2,Y)$ under the standard model
can only be in the massless string spectrum if the following condition is
verified (see e.g., \FIQ )
\eqn\kac{
{{C(R_3)}\over {k_3+3}}\ +\ {{C(R_2)}\over {k_2+2}}\ +\
{{Y^2}\over {k_1}}\ \leq \ 1
}
where $C(R_i)$ is the quadratic Casimir of the $R_i$ representation
of the group $SU(i)$ and the usual asignements for the hypercharges
has been assumed (e.g., $Y(Q_L)=1/6$). Now, in order for the
$e_R$ to be in the massless spectrum one necessarily has $k_1\geq 1$
\SCHEL .
This is the only model independent constraint which is known about
$k_1$. The hypercharge generator in string models (see e.g. Appendix C in ref.
\FENO ) can be written
in terms of the 16 Cartan subalgebra bosonic coordinates $X_I$ of
$E_8\times E_8$ as $Y=i\sum _IQ_I\partial X_I$. One then
finds $k_1=2\sum _IQ_I^2$. Since in each model the hypercharge
is embedded into $E_8\times E_8$ in a different way (i.e.,
one has different $Q_Is$), the obtained results for $k_1$
is different and can be only computed in a model by model
basis.
Concerning the actual values of $k_1$ in specific models,
in the orbifold examples constructed up to now the value of
$k_1$ is never the canonical 5/3 and has in fact the tendency to be
larger. Other models are specifically constructed in order to get
a value $k_1=5/3$. As discussed above, values for $k_1$ slightly
smaller  like $3/2$ or $7/5$ would be preferable from the
point of view of direct standard model unification into a string.
It would be very important to look for model independent
information about $k_1$. In the case of GUTs it is
 possible to obtain a variety of values for $k_1$ which
turn out to be always bigger than 5/3 if one embeds the standard
model into a bigger simple group. I do not know of
any equivalent statement concerning string models.

It would be interesting to find specific string examples with
the preferred $k_1$ values in the range 1.4-1.5. On the other
hand, the equivalent statement for SUSY-GUTs is that it would be
interesting to find GUTs in which the natural value of $M_X$
is of order $10^{16}$ GeV. Furthermore, in the GUT case one
would also have to find a natural mechanism for doublet-triplet
splitting. Which of both alternatives one should take is at the moment a
question of taste. However, it is fair to say that leaving
$k_1$ as free parameter (as one should do till we have a general
handle on it) allows for gauge coupling unification
(within expected uncertainties) at the
string scale.

\

\bigskip

\bigskip

\bigskip

\bigskip\nobreak
\centerline{Figure  }
\bigskip
\noindent \vbox{\baselineskip 14pt \noindent
Constraints in the $sin^2\theta _W(M_Z)-\alpha _s(M_Z)$ plane
coming from unification of gauge coupling constants}

\vskip8pt

\listrefs

\bye